\documentclass[useAMS,usenatbib]{mn2e}
\usepackage{amsmath}
\usepackage{wasysym}
\usepackage{graphicx}
\usepackage{epsfig}
\usepackage{epic}
\usepackage{eepic}
\usepackage{color}
\usepackage[usenames,dvipsnames]{xcolor}
\usepackage{float}

\newcommand{\apjl}{ApJL}                                                       
\newcommand{\aap}{A\&A}

\newcommand{\apjs}{ApJS}

\def\H0m{\mbox{$\langle H_0 \rangle$}}
\def\vpec{\mbox{$\vec{v}_{\rm p}$}}

\def\LCDM{\mbox{$\Lambda$CDM}}

\def\Mpch{\mbox{$\,h^{-1}$Mpc}}

\def\M200{\mbox{$M_{\rm 200 }$}}
\def\Msunh{\mbox{$h^{-1}M_\odot$ }}

\def\R200{\mbox{$R_{\rm 200 }$}}

\def\V200{\mbox{$V_{\rm 200 }$}}
\def\Nbody{\mbox{$N$-body }}

\def\div{\mbox{$\nabla\cdot$}}

\def\dV{\mbox{${\rm d}V$}}

\newcommand{\kms}{\mbox{\,s$^{-1}$\,km}}
\newcommand{\kmsmpc}{\mbox{\,s$^{-1}$\,km\,Mpc$^{-1}$}}

\newcommand{\mbi}[1]{\mbox{\boldmath$#1$}}
\renewcommand{\vec}[1]{\mbox{\boldmath$#1$}}
\newcommand{\eq}[1]{\mbox{Eq.~\ref{#1}}}

\newcommand{\lsim}{\mbox{${\,\hbox{\hbox{$ < $}\kern -0.8em \lower 1.0ex\hbox{$\sim$}}\,}$}}
\newcommand{\gsim}{\mbox{${\,\hbox{\hbox{$ > $}\kern -0.8em \lower 1.0ex\hbox{$\sim$}}\,}$}}

\newcommand{\shortminus}{\scalebox{0.5}[1.0]{\( - \)}}

\def\beqn{\vspace{2mm}
\begin{eqnarray}} 
\def\eeqn{\vspaceg{2mm} 
\end{eqnarray}}

 \def\la{\langle}

\newcommand{\be}{\begin{equation}}
\newcommand{\ee}{\end{equation}}
\newcommand{\ba}{\begin{eqnarray}}
\newcommand{\ea}{\end{eqnarray}}
\newcommand{\brr}{\begin{array}}
 
\newcommand{\err}{\end{array}}
\newcommand{\bc}{\begin{center}}
\newcommand{\ec}{\end{center}}



\def\dHcosmo{\mbox{$-1.30 \pm 0.24$ }} 
\def\dHcosmob{\mbox{$-1.76 \pm 0.21$ }} 

\def\moatinner{\mbox{$27\Mpch$} }
\def\moatouter{\mbox{$47\Mpch$} }

\title[Cosmic flows and the expansion of the Local Universe]{Cosmic flows and the expansion of the Local Universe\\ from nonlinear phase-space reconstructions}

\author[Steffen He{\ss} and Francisco-Shu Kitaura]{
Steffen He{\ss} \thanks{E-mail: hess.aip@gmail.com} \& Francisco-Shu Kitaura
\thanks{E-mail: kitaura@aip.de, Karl-Schwarzschild fellow}\\ 
Leibniz Institute for Astrophysics, An der Sternwarte 16,
14482 Potsdam, Germany \vspace{0.2cm}
}
\begin{document}


\maketitle
\label{firstpage}
\begin{abstract}
In this work we investigate the impact of cosmic flows and density perturbations on Hubble constant $H_0$ measurements using nonlinear phase-space reconstructions of the Local Universe.  In particular, we rely on a set of 25 precise constrained \Nbody simulations based on Bayesian initial conditions reconstructions of the Local Universe (LU) using the 2MRS galaxy sample within distances of about 90 {\Mpch}. These have been randomly extended up to volumes enclosing distances of 360 {\Mpch}  with augmented Lagrangian perturbation theory (750 simulations in total), accounting in this way for gravitational mode coupling from larger scales, correcting for periodic boundary effects, and estimating systematics of  missing attractors ($\sigma_{\rm large}=134$ $\kms$ ).
We report on Local Group (LG) speed reconstructions, which for the first time are compatible with those derived from CMB-dipole measurements: $|v_{\rm LG}|=685\pm137$ $\kms$.  The direction $(l,b)=(260.5\pm 13.3,39.1\pm 10.4)^\circ$ is found to be compatible with the observations after considering the variance of large scales.

Considering this effect of large scales, our local bulk flow estimations assuming a $\Lambda$CDM model are compatible with the most recent estimates based on velocity data derived from the Tully-Fisher relation.

We focus on low redshift supernova measurements out to $0.01<z<0.025$, which have been found to disagree with probes at larger distances. Our analysis  indicates that there are two effects related to cosmic variance contributing to this tension. The first one is caused by the anisotropic distribution of supernovae, which aligns with the velocity dipole and hence induces a systematic boost in $H_0$. The second one is due to the inhomogeneous matter fluctuations in the Local Universe. In particular a divergent region  surrounding the Virgo Supercluster is responsible for an additional positive bias in $H_0$. Taking these effects into account yields a correction of $\Delta H_0=\dHcosmob {\rm \kmsmpc}$, thereby reducing the tension between local probes and more distant probes. Effectively $H_0$ is lower by about 2\%.
\end{abstract}


\section{Introduction}
\label{Introduction}

Measurements of the Hubble constant coming from different probes show high variations. In particular cosmic microwave background (CMB) measurements reported by the \citet{Planck2013} and recession velocity measurements of Type Ia Supernovae (SNe Ia) \citep[see e.g.,][]{Riess2009,Perlmutter1999} show a discrepancy of about $2.4\sigma$. 
But there is also evidence for discrepancies between more local probes.
\citet{Jha2007} reported $6.5 \pm 1.8\%$ higher values for a local sample ($z<0.025$ corresponding to about $<75 \Mpch$), as compared to the more distant set.

There are presumably two combined main effects which could explain this discrepancy. First, cosmic variance (CV) of the peculiar velocity distribution can affect the measurement. Most notable is the effect of monopoles in the velocity field or ``Hubble bubbles'' such as described by 
(\citet{Zehavi1998,Jha2007,Conley2007}.
Second, there may be systematic errors in the standardisation of SNe Ia due to environmental dependencies \citep[][]{Rigault2013,Rigault2014}.

A number of studies have focused on estimating the influence of CV from large sets of simulations (see e.g. \citet{Marra2013,Wojtak2014,BenDayan2014}), as local probes are particularly sensitive to the local density and velocity field.
These theoretical studies analyse the impact of random seeded mock catalogues on the Hubble measurement, thereby studying cosmic variance in a probabilistic way \citep[see also][]{Hui2006,Davis2011}.

As an alternative to statistical CV estimates, one can use the actual velocity information of the Local Universe to reduce the influence of CV on $H_0$ estimations. 
Interestingly, \citet{Wiltshire2013} find from radial velocity data, that the amplitude over the Hubble flow changes markedly over the range 32 to 60 $\Mpch$. 
\citet[][]{Riess1997} connected Galaxy surveys with the velocity field from SN.
\citet[][]{Neill2007} and \citet{Riess2009} managed to find moderate corrections to the Hubble constant estimate based on peculiar velocity corrections derived from galaxy redshift data using linear theory. 

We want to extend these works by investigating the inferred  nonlinear  velocity field from high precision constrained N-body simulations.
The simulations are based on Bayesian self-consistent phase-space reconstructions of the Local Universe as measured by the Two-Micron Redshift Survey (2MRS) galaxy catalogue \citep{Huchra12}. In particular, using the nonlinear velocity field we will study its influence on low redshift supernovae measurements ($z<0.025$) \citep[][]{Hess2013}.

This paper is structured as follows. First, we study  in  \S\ref{sec:divergence} the qualitative impact of cosmic variance through peculiar motions on the estimation of the Hubble constant from a theoretical perspective. In \S\ref{sec:cosmicflows} we compute the peculiar velocities from constrained simulations of phase-space reconstructions. The resulting shifts on the Hubble constant measurement are presented in \S\ref{sec:Results}. Then we discuss our results in \S\ref{sec:dis}. Finally, we present our conclusions.

\section{Relation between the Hubble constant and the velocity field}
\label{sec:divergence}
Let us consider the Local Volume, where the redshift evolution is negligible.
The radial recession or Hubble flow is then given by $\vec{v_{H}} \equiv H_0\vec{r}$, with $\mbi r$ being the line-of-sight distance vector, and $H_0$ the Hubble constant at redshift zero.
 Hence, the Hubble constant  $H_0$  measuring the expansion speed of the Universe is directly related to the divergence of the Hubble flow by
\begin{equation}
  \label{eq:hubblediv}
  H_0= \frac{1}{3} \div \vec{v_{H}}\,.
\end{equation}
The observed Hubble constant $\H0m_{\rm obs}$ is in general measured from an average of $N$ discrete distance tracers 
\begin{equation}
  \label{eq:hubble}
  \H0m_{\rm obs} \equiv  \frac{1}{N} \sum_{i=1}^N H_0^i \equiv \frac{1}{N} \sum_{i=1}^N  \left( \frac{\vec{v}_i \cdot
\hat{\vec{r}}_i}{{r}_i}\right)\,,
\end{equation}
where each tracer $i$ contributes to the average with $H_0^i$, $\hat{\vec{r}}$ is the unitary line of sight vector, and the velocities are given by the sum of the Hubble flow and the peculiar motion term
\begin{equation}
  \label{eq:vsplit}
\vec{v}=H_0\vec{r}+\mbi v_{\rm p}\,,
\end{equation}
 with $\mbi v_{\rm p}$ being the peculiar velocity of each tracer (we refer to \S~\ref{sec:Distancecalculation} for a relativistic treatment).
 Each tracer $i$ at a distance $r_i$ approximately measures the Hubble constant within the volume  $V_i=\frac{4\pi}{3}r_i^3$. However, this will in general only be true under the assumption that the line of sight projected peculiar velocities are isotropic, as can be seen from applying Gauss theorem to the volume averaged Hubble constant
\begin{eqnarray}
  \label{eq:hubble_gauss}
  \H0m_{V_i }
     &\equiv& \frac{1}{3 V_i} \int_{V_i}{\rm d}V \, \div \vec{v}= \frac{1}{3 V_i} \oint_{\partial V_i} {\rm d}\vec{S} \cdot \vec{v} \ \\
\label{eq:int_shell}
     &=& \frac{1}{4\pi} \oint_{\partial V_i}{\rm d}\theta {\rm d}\phi\,  {\rm sin}\phi \,\frac{\vec{v} \cdot \vec{\hat{r}}}{r_i}\,.
\end{eqnarray}
 
Hence, an anisotropic line of sight peculiar velocity field will introduce a systematic bias in the Hubble constant measurement. This effect is amplified by an anisotropic tracer distribution in the sum of Eq.~\ref{eq:hubble}, as we will show below.

  Unfortunately, supernovae and their host galaxies, are not evenly distributed in space, but are affected by various radial selection effects and a complex biasing w.r.t. the underlying dark matter distribution. Let us therefore consider a simple model, including a radial selection function depending on the distance $r$ and a bias as a function of the local density field $\rho=\rho(\vec{r})$ \citep[for more complex non-local biasing relations see, e.g., ][]{Mcdonald09}
\begin{equation}
  \label{eq:SN_prob}
  f = f\left(r,\rho(\vec{r}) \right)\,.
\end{equation}
We can now obtain  a more realistic model of the  observed Hubble constant by introducing the selection biased  Hubble constant   $H_0^{f(r_i,\rho(\vec{r_i}))}\simeq H_0^i$ and $ \langle H_0^f \rangle= \langle H_0^{f(r_i,\rho(\vec{r_i}))} \rangle$ defined as
\begin{eqnarray}
  \label{eq:hubble_split}
   \langle H_0^f  \rangle &\equiv&   \langle \frac{(f\vec{v})\cdot \hat{\vec{r}}}{{\langle f \rangle r}} \rangle 
= H_0+\frac{1}{3V_f} \int_{V}\dV \div\left(f \vec{v}_{\rm p} \right)\,,	
\end{eqnarray}
where we have introduced the effective volume  $V_f = \int \dV \,f(\vec{r})$ and the average $\langle f \rangle = \frac{1}{V}\int \dV \,f = \frac{V_f}{V}$. Furthermore we used  \eq{eq:vsplit} and have applied Gauss theorem in the reversed order to Eq.~\ref{eq:hubble_gauss}.

We find the deviation between the true value of the Hubble constant and the one obtained from a selection biased tracer is given by
\begin{eqnarray}
  \label{eq:hubble_split2}
\lefteqn{ H_0 -\langle H^f_0 \rangle  =  -\frac{1}{3V_f}  \int_{V}\dV  \div(f\vec{v}_{\rm p})} \\
  \label{eq:hubble_split_terms}
&&\hspace{0cm} = -\frac{1}{3V_f} \int_{V} \dV  \left( f \div\vec{v}_{\rm p}  + \vec{v}_{\rm p}\cdot\vec{\hat{r}} \frac{\partial f}{\partial r}+\vec{v}_{\rm p} \cdot \nabla \rho \frac{\partial f}{\partial \rho}\right) \,.
\end{eqnarray}

It is obvious from this equation that in the limiting case of negligible peculiar motions the difference vanishes. However, interestingly, there are more contributions in the general case, beyond the effective divergent flow  term (first term in Eq.~\ref{eq:hubble_split_terms}).
Local peculiar velocities modulate terms dominated by radial selection  effects and density perturbations (second and third terms, respectively). 

 We conclude from this analysis that a proper measurement of the Hubble constant should include accurate  peculiar velocity flow corrections under the consideration of the distribution of matter beyond divergent flow corrections, as we will show in this work.

\begin{figure}
\vspace{-2mm}
\centering
\hspace{-1mm}
\includegraphics[width=.48\textwidth]
{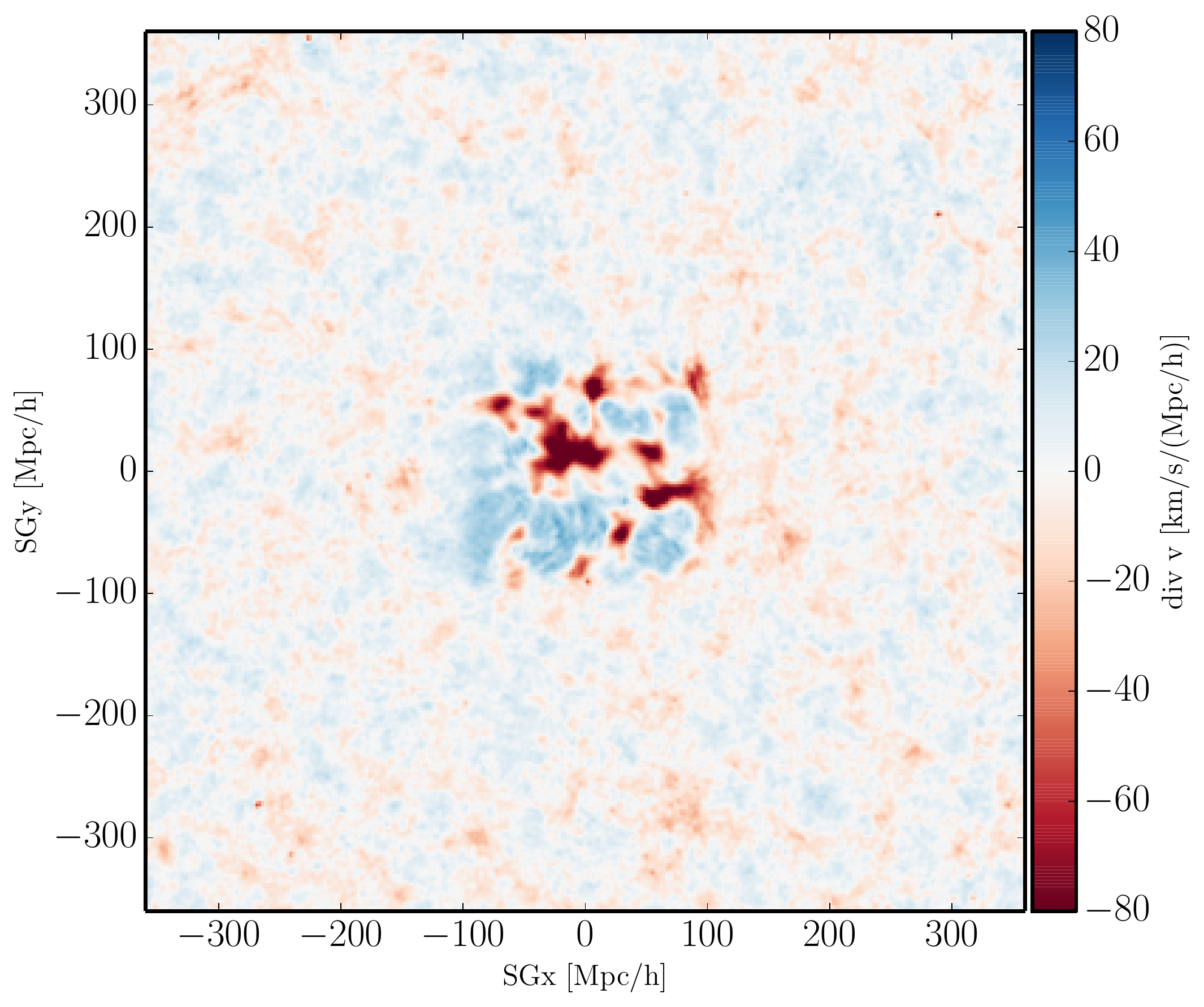}
\vspace{-4mm}
\caption{Supergalactic (SG) XY slice, showing the velocity divergence through the averaged augmented constrained simulations. Each of the 25 constrained simulations has been rerun $30$ times with random large scale modes in a box of $720\Mpch$ and averaged. Additionally a Gaussian smoothing of $\sigma=2.8\Mpch$ has been employed. The colorscale shows collapsing regions in red and diverging regions in blue. We note that the velocity divergence is not strongly affected by large scale modes and hence the features at the edge of the original constrained $180\Mpch$ box remain visible after averaging.
}
\label{fig_mapsfull}
\end{figure}

\section{Cosmic flows reconstruction}
\label{sec:cosmicflows}

The supernova data permit one to estimate the Hubble constant by linear regression of their redshifts against their luminosity distances.
Their measured redshift positions are however affected by peculiar motions of the large scale structure. Therefore a correction of the redshift space distortions is necessary to improve the Hubble constant estimation, as discussed in the previous section.

In this section we describe our nonlinear peculiar halo velocity field reconstruction, followed by the study of the influence of larger scales beyond the reconstructed volume and the dependence on cosmological parameters. Finally we  show the supernova peculiar motion correction based on simulations using the halo phase-space reconstruction.

\subsection{The halo peculiar velocity field from nonlinear phase-space reconstructions}
\label{sec:Constrained simulations}

We aim at obtaining a full nonlinear reconstruction of the peculiar velocity field.
To solve this problem we rely on the \textsc{kigen}-code \citep[][]{Kitaura_kigen,Kitaura_2mrs}, updated with a number of improvements reported in \citet[][]{Hess2013} and \citet[][]{Nuza2014}.
This  method is based on a Bayesian networks machine learning algorithm, which iteratively samples Gaussian fields, whose phases are constrained by the distribution of observed tracers given a structure formation and a cosmological model. 

In particular, it employs  two Gibbs-sampling steps. In the first step, the phases of the Gaussian fields are sampled given a distribution of matter tracers of proto-haloes at Lagrangian initial conditions, for which a lognormal-Poisson distribution \citep[][]{Kitaura_log} including a linear Lagrangian bias  (power law bias within the lognormal framework) is assumed \citep[see also][]{patchy}.
The log-normal model is a fair assumption in a Lagrangian co-moving description when shell-crossing is negligible \citep[see][]{Coles1991,Kitaura_lin}.

In the second step, the positions of the proto-halo tracers are obtained given the initial Gaussian phases from the previous step. Here a Chi-squared likelihood comparison minimising the quadratic distance between the final modelled matter positions and the observed positions of galaxies is used.   The structure formation model including redshift space distortions connecting the initial Gaussian field with the observations  
is based on augmented Lagrangian perturbation theory  \citep[ALPT,][]{Kitaura_ALPT}. An uncertainty variance in the Chi-square of 1{\Mpch} accounts for the inaccuracy of the ALPT approximation and the probability that a galaxy is associated to a particular halo.

This scheme is iterated until it reaches convergence of the power spectra of the initial Gaussian fields and the cross-power spectra between the reconstructed dark matter field in redshift space and the observed galaxy field. Then a couple of thousand additional iterations are run to produce an ensemble of initial conditions compatible with the observations.
The  Lagrangian bias  is self-consistently constrained to yield unbiased power spectra with respect to the theoretical linear power spectrum model. 

The method relies on the position measurements of 31017 galaxies from the Two-Micron Redshift Survey (2MRS) galaxy catalogue \citep{Huchra12} as tracers of the matter distribution within a cubic box of $180\Mpch$. Based on this measured matter density field, we have selected the set with the highest correlation between the measured and the simulated density fields to perform 25 constrained \Nbody  simulations \citep{Hess2013}. These comprise a cubic volume of $180 \Mpch$ and simulate the nonlinear structure formation assuming WMAP7 \citep[][]{wmap7} cosmology. 
 These simulations resemble the Local universe on scales larger than $2 \Mpch$ and resolve spherical overdensity halos 
with masses $m>1.6\times10^{11}\Msunh$. Therefore most of the heavily star-forming and hence supernova-forming objects are resolved.

This permits us to study the full nonlinear peculiar velocity field based on the halo population, and supposes a considerable improvement with respect to the dark matter peculiar velocity field based on second order Lagrangian perturbation theory presented in \citet[][]{Kitaura_2mrs}.

A cosmic web study performed with the same simulations in \citet[][]{Nuza2014}
found that we have a closely fair sample at scales of about $60 \Mpch$.
To analyse flow and density patterns on scales of our reconstructed volumes 
of about  $90 \Mpch$ we also need to consider the influence of  inhomogeneities of larger scales.

\begin{figure*}
\vspace{-1cm}
\begin{tabular}{cc}
\includegraphics[width=.48\textwidth]{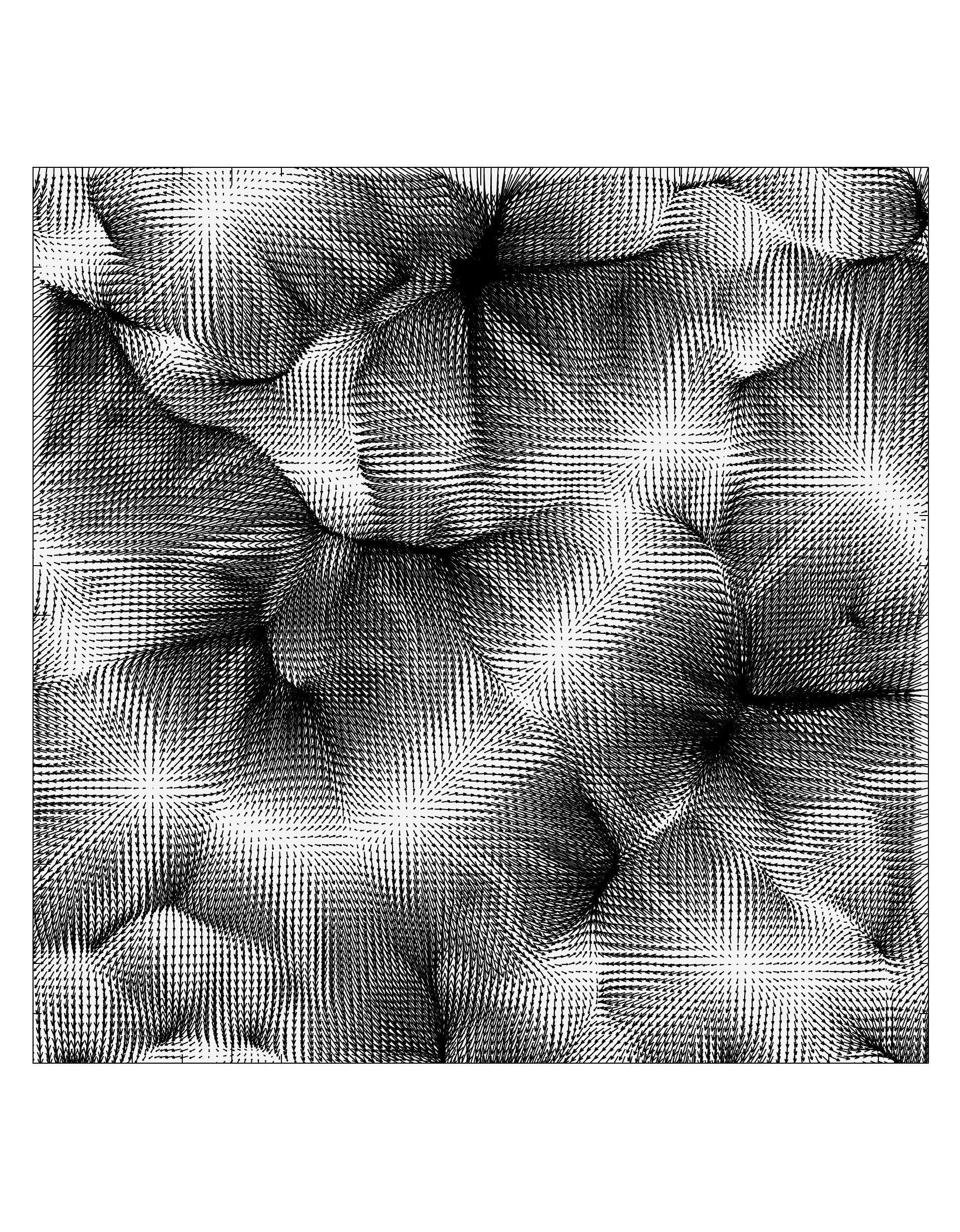}
\put(-260,152){\rotatebox[]{90}{SGy [Mpc/h]}}
\put(-246,215){$50$}
\put(-243,153){$0$}
\put(-247,90){$\shortminus50$}
\put(-140,22){\rotatebox[]{0}{SGx [Mpc/h]}}
\put(-192,33){$\shortminus50$}
\put(-123,33){$0$}
\put(-63,33){$50$}
\includegraphics[width=.48\textwidth]{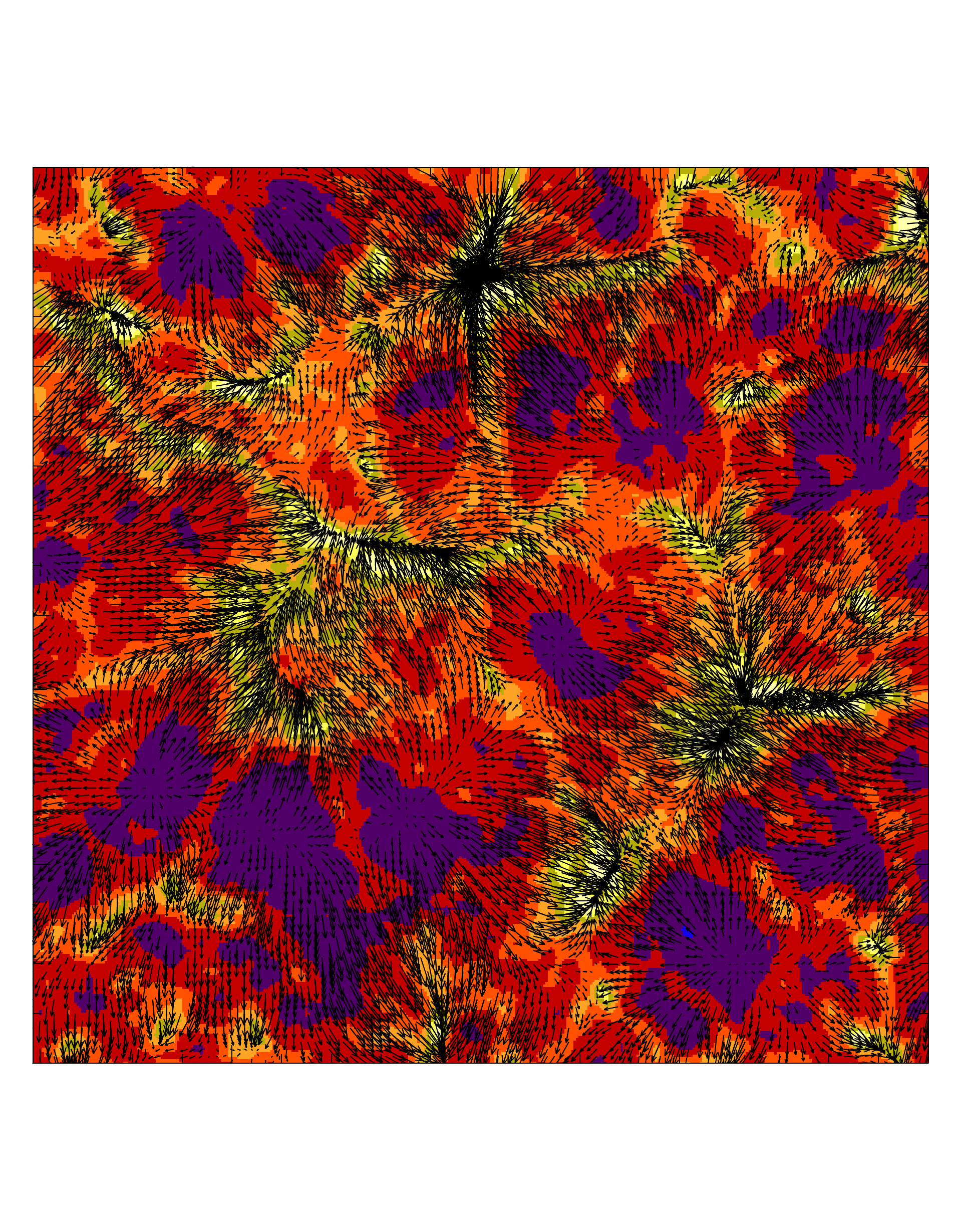}
\put(-140,22){\rotatebox[]{0}{SGx [Mpc/h]}}
\put(-192,33){$\shortminus50$}
\put(-123,33){$0$}
\put(-63,33){$50$}
\end{tabular}
\vspace{-0.7cm}
\caption{On the left: slices of the averaged $v_x$-$v_y$ velocity field from the ensemble of constrained simulations, on the right: including the underlying dark matter density field on a grid with cells of 1.4 $h^{-1}$\,Mpc side. For visualisation purposes of the shell crossing regions of the peculiar velocity field coinciding with high density regions, the dark-blue, the reddish, and the light-yellow colour-codes indicate low, close to mean, and high densities, respectively. The length of the arrows is proportional to the average speed at that location. 
}
\label{fig_flows}
\end{figure*}
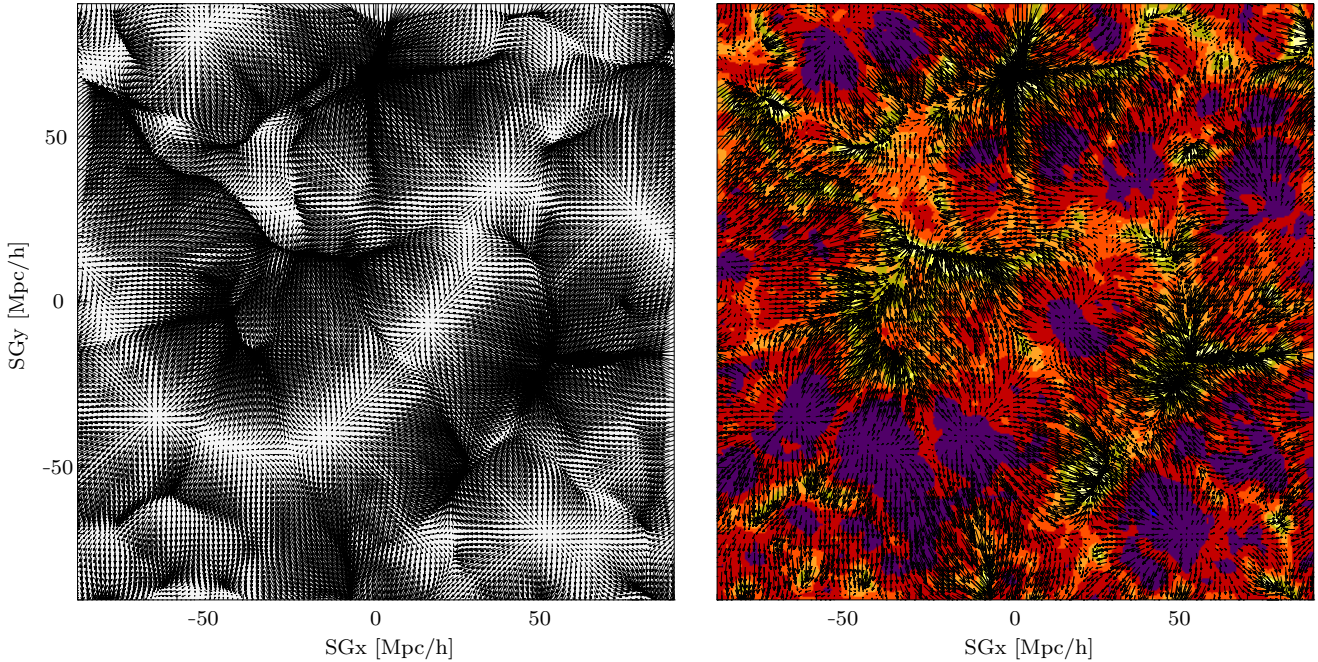

\begin{table}
\rotatebox[]{0}{
\begin{tabular}{|c|c|c|} 
& { ($l,b$)$^\circ$}&{ $|v|$ [$\kms$]}\\\hline
{$v_{\rm CMB}$} & $(276\pm3,30\pm3)$ &$627\pm22$ \\ 
{$v_{\rm LG}$} & $(260.5\pm2.5,39.2\pm1.7)$ &$685\pm36$\\  
{$v^{\rm large}_{\rm LG}$} & $(260.5\pm13.3.5,39.2\pm10.4)$ &$685\pm137$
\end{tabular}
}
\caption{ \label{tab:v}  Local group velocity from CMB-dipole measurements $v_{\rm CMB}$ (Kogut et al. 1993), from the constrained \Nbody simulations within $2.3 {\Mpch}$ distance $v_{\rm LG}$, and including large scale modes uncertainty estimates: $v^{\rm large}_{\rm LG}$.}
\end{table}

\subsection{Influence of large scales}
\label{sec:Influence of large scales}

The constrained simulations only probe distances of \mbox{$\sim 90 \Mpch$} along the SG axis (and of $\sim 127\Mpch$  along the diagonal within the SG XY plane). In this way the Shapley concentration (SC), with its centre located at distances of about 130-140 Mpc/h (in the diagonal direction within the SG plane), is mainly excluded from our volume  (see for instance Fig.~19 in \citet[][]{Erdogdu2006} including the SC). We note that the reconstructed volume was limited to distances where uncertainties due to the particular selection function, especially the Kaiser-rocket effect, are small \citep[][]{Nusser2014}. Therefore missing attractors, such as the SC,  and larger modes are expected to have an effect on our reconstructions  \citep[see e.g.,][]{Nusser2014}. Another source for systematic errors comes from the periodicity assumption within the reconstruction process. Let us represent the resulting halo peculiar velocity field from our constrained \Nbody simulations by $\mbi v_{\rm h}(\mbi x)$, where $\mbi x$ represents Eulerian real-space.
To asses the influence of the above mentioned effects we embed the constrained initial conditions in a bigger volume of $(720\Mpch)^3$, effectively including distances of up to $360\Mpch$ with respect to the Local Group. We exploit the advantage of having reconstructed the initial conditions, partially following the methods by  \citet[][]{Tormen1996} and \citet{Schneider2011}. In particular we compute the white noise (initial density field divided by the square root of the power spectrum) of the set corresponding to the 25 best reconstructed fields and augment it by adding random phases of unity variance beyond the reconstructed volume to boxes of $720\Mpch$ side (similar to \citet{Hoffman2001}). We finally multiply the augmented white noise with the square root of the power spectrum in Fourier space to produce the  primordial Gaussian density field from which we compute the peculiar velocity field with ALPT. 

To accurately estimate the effects of mode-coupling beyond the reconstructed volume,  we compute the differences between the constrained and augmented boxes in Lagrangian space (we denote Lagrangian coordinates with $\mbi q)$. For definiteness we compute difference using ALPT for both boxes. We can therefore define the correction, as the difference between the long range (enclosing the big volume) and the constrained component (enclosing the small volume): 
\begin{equation}
  \label{eq:vdiff_lagr}
  \Delta \mbi v_{\rm ALPT}^{\rm large}(\mbi q)=\mbi v^{\rm large}_{\rm ALPT}(\mbi q)-\mbi v_{\rm ALPT}(\mbi q)\,.
\end{equation}
We then use the constrained displacement field to evaluate the correction in Eulerian-space 
with $\mbi x=\mbi q+\mbi \Psi$, where $\mbi \Psi$ is given by the smaller constrained volume. In  this way we do not alter the displacement fields, which have been accurately constrained within the self-consistent reconstruction process. We can finally compute the long range corrected halo peculiar motions by adding the correction to the full nonlinear constrained component:
\begin{equation}
  \label{eq:vdidff_eul}
\mbi v_{\rm h}^{\rm large}(\mbi x)\simeq\mbi v_{\rm h}(\mbi x)+\Delta \mbi v^{\rm large}_{\rm ALPT}(\mbi x)\,.
\end{equation}
For each of the 25 constrained small boxes we compute the mean and variance of 30 augmented ALPT simulations comprising 750 realisations in total.
If not denoted otherwise we compute averages on these 750 realisations.  Fig.~\ref{fig_mapsfull} shows the influence of modes beyond the reconstructed volume on the divergence of the velocity field in one of these augmented ALPT simulations.

From this analysis we get the following results.
\begin{enumerate}
\item By inspecting the ensemble mean of the realisations we find $\langle \mbi v_{\rm h}^{\rm large}(\mbi x)\rangle$, a systematic deviation of $39 \kms$ in the direction of ($l,b$) = $(37, 19)^{\circ}$ respectively, caused by the periodicity assumption or the missing modes, since the attractors beyond the reconstructed volume cancel out in the ensemble average. An illustration of the cosmic velocity field in the Local Universe derived from our calculations is shown in Fig.~\ref{fig_flows}. The caustics of the peculiar velocity field show a remarkable correlation with well-known structures like the Local Super-cluster, the Great Attractor, the Coma, and the Perseus-Pisces  clusters, indicating the accuracy of the peculiar velocity field.
Let us report, as part of the results of this work, the peculiar motion of the Local Group with a speed of $|v_{\rm LG}|=685\pm36$ {\kms} and pointing towards galactic longitude and latitude of $(l,b)=(260.5\pm2.5,39.2\pm1.7)^\circ$, respectively, after taking into account the mean correction. We have considered the mean peculiar velocity of LG like haloes contained within 2.3 {\Mpch} distance to the 
location of the observer. This distance accounts for the $2-3$ {\Mpch} uncertainty in the location of haloes \citep[][]{Hess2013}.
 Here, one should note that the standard error represents the error of the mean of the ensemble of constrained halo catalogues, and a number of systematic effects are not included, as we will do below.
We find however, already an interesting result with respect to previous works based on linear or even second order Lagrangian perturbation theory, as the speed is considerably larger than in those previous studies \citep[see e.g.][]{Kitaura_2mrs}. This speed is compatible with independent CMB-dipole measurements ($|v_{\rm LG}|=627\pm22$ {\kms} \citep[][]{Kogut1993}). 

\item We find by inspecting the variance of the realisations $\sigma_{\rm large}\equiv\sqrt{\langle \mbi v_{\rm h}^{\rm large}(\mbi x)^2-\langle \mbi v_{\rm h}^{\rm large}(\mbi x)\rangle^2\rangle}$, that the whole re-simulated $180 \Mpch$ box typically moves with  $\sigma_{\rm large}= 134 {\kms}$ given by the standard deviation. 
This is not in opposition with the framework of observations considering data on intermediate (i.e. \citet{Courtois_flows_2012,Ma2014}) and larger volumes (see the bulk flow estimates of distances $>90 \Mpch$  in \citet{Watkins2009,Feldman2010,Colin2011,Feindt2013}).
\end{enumerate}
By including these uncertainty estimates we find compatible results with the CMB-dipole local group velocity measurement as summarised in  Tab.~\ref{tab:v}.
However, our analysis does not specify the direction of the missing bulk motion of the inner $180 \Mpch$ box. We can compare the direction with the CMB-dipole information (
$|v_{\rm LG}| = 627\pm22 \kms$ towards ($l,b$) = $(276\pm3, 30\pm3)^{\circ}$ \citet[][]{Kogut1993}).

To reconcile the halo velocity in the centre of the box with the Local Group velocity we could assume a bulk motion of the simulation box of $|v^{\rm large}| = 187\kms$ towards ($l,b$) = $(354,-39)^{\circ}$ which is consistent with our analysis, corresponding to  $1.4\sigma_{\rm large}$. Interestingly, this missing component is closely perpendicular to the constrained flow speed, which was already compatible with the CMB measurement. However, even though reproducing the CMB dipole in the center, the box motion depends to no small extend on the original velocity of the box center, which is highly non-linear and little constrained. Therefore it is not a reliable proxy for the box motion.

\begin{figure}
\centering
\includegraphics[width=.47\textwidth]
{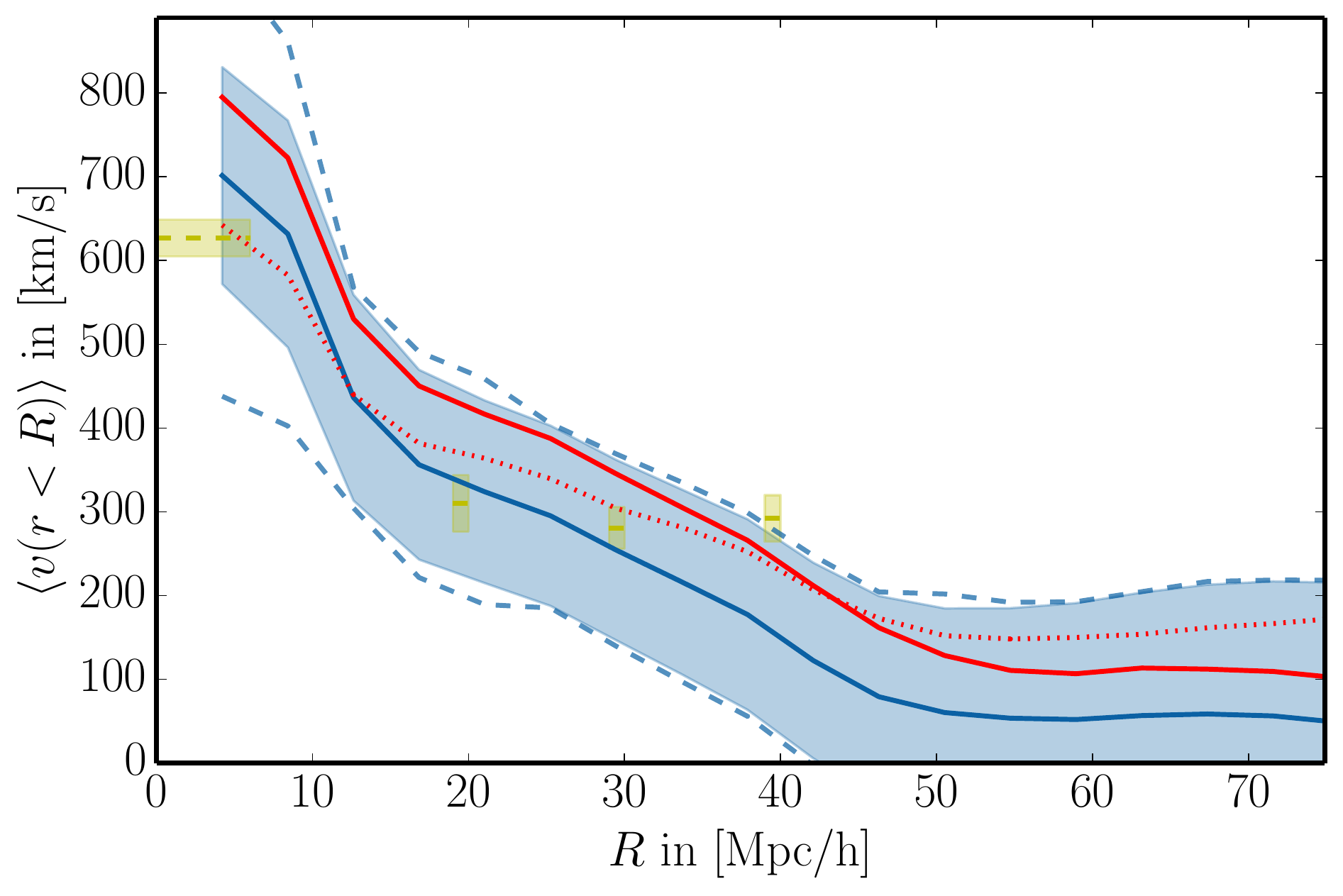}
\vspace{-1.5mm}
\caption{Average bulk flow within radius $R$ of potential SN hosting halos from the ensemble of constrained simulations. Shown are estimates without (blue) and with restframe corrections of the simulation box following either the assumption that the box center (red dotted) or the box itself (red solid) moves towards the CMB dipole.
The blue shaded region marks the uncertainty containing the error of the ensemble mean and the standard deviation of the bulk flow estimate. The dashed lines extend the uncertainty due to differences in cosmology. As comparison serves the Local Group speed (Kogut et al. 1993) and TF measurements (Hong et al. 2014) in yellow.
}
\label{fig:bulk}
\end{figure}

Hence, to account for super box scales, we assume a box motion of the box according to $\sigma_{\rm large}$ in the direction of the CMB dipole. This is compatible with \citet{Watkins2009,Feldman2010,Colin2011,Feindt2013} who indeed found that on scales of at least $\sim 100 \Mpch$ the bulk flow is short in amplitude w.r.t. CMB dipole but in constant direction.
Fig.~\ref{fig:bulk} shows the average velocity of LG like haloes with and without the large scale correction. For definiteness it shows the ensemble average of halos enclosed on increasing radii $\langle \mbi v_{\rm h}^{\rm large}(r<R)\rangle$. 

There has been a controversy on the amplitude of local bulk flows between different studies \citep[see e.g.][]{Watkins2009,Feldman2010,Colin2011,Turnbull2012,Feindt2013,Nusser2014}. Especially \citet{Kashlinsky2008} and \citet{Watkins2009}  reported  bulk flows which are too high to be compatible with standard \LCDM-cosmology.  \citet{NusserDavis2011} estimated a bulk flow of $333 \pm 38 \kms$ at $40 \Mpch$ using the SFI++ Tully-Fisher sample. \citet{Turnbull2012} derived a bulk flow of $249 \pm 76 \kms$. However, many of these studies use
varying assumptions, volumes and sky coverages which makes a direct 
comparison challenging.
We use the recent full sky study of \citet{Hong2014}, who used the Tully-Fisher relation with the 2MTF galaxies and estimated bulk flows of $310.9 \pm 33.9 \kms$, $280.8 \pm 25.0 \kms$, and $292.3 \pm 27.8 \kms$ at depths of $20 \Mpch$, $30 \Mpch$ and $40 \Mpch$ and find our estimations to be consistent (see Fig.~\ref{fig:bulk}).

\subsection{Cosmology dependence}

The reconstruction scheme as well as the constrained simulations have been performed assuming WMAP7 cosmology. However both the reconstruction and the simulations have been performed in units of $h^{-1}$, without choosing a particular $H_0$, and are hence suitable for this study. Other cosmological parameters have an influence on the peculiar velocity field. To gauge this effect we have performed a constrained simulation with Planck cosmology \citep[][]{Planck2013} using existing constrained initial conditions. We find that halo velocities tend to be slightly higher especially at small scales. In Fig.~\ref{fig:bulk} we indicate  the uncertainty introduced by the difference of the two cosmologies (dashed lines). We note that this study tends to overestimate the effects of the cosmological parameters. The reason being that the initial conditions have been found with WMAP7 cosmology. A self-consistent reconstruction of the initial conditions with PLANCK cosmology would certainly yield closer results.  Nevertheless, the bulk flows remain very similar at radii larger than about 10 $h^{-1}$ Mpc, demonstrating the robustness of our results.

\subsection{Supernova peculiar motion correction}
\label{sec:Identifying peculiar velocities}

In the previous sections we have described the method used to obtain an ensemble of  halo distributions with the corresponding peculiar motions constrained on the Local Universe, including an accurate assessment of uncertainties due to systematics (periodic boundary conditions and missing modes from larger scales) and cosmic variance (missing attractors beyond the reconstructed volume).
The ensemble of catalogues permits us to account for uncertainties derived from the reconstruction method itself. We note that the phase space reconstructions reproduce information of the large-scale structure down to about 2{\Mpch}, meaning that structures below these scales are essentially random (see \citet{Hess2013}). This is reflected in the lack of correlation on these scales between realisations within the ensemble.

To each SN we assign the mean of peculiar velocities of constrained haloes ($\mbi v_{\rm h}^{\rm large}$) which are nearby in redshift space \citep[see][]{Nuza2014}.

Let us call the resulting estimated supernova peculiar motion $\mbi v_{\rm SN}$.
In this way the information of the ensemble of solutions is used to obtain a single and conservative correction. We note that we are neglecting the peculiar motion of supernovae within galaxies. 

We find that the amplitudes of the peculiar velocities tend to get slightly underestimated due to the conservative nature of the ensemble mean. The distribution of radial velocities for the reference haloes have $\sigma_{vH}=282 {\kms}$, whereas the radial velocities assigned to the supernovae have  $\sigma_{v{\rm SN}}=224 {\kms}$. Given the additional uncertainty of the assignment itself, of $\sigma_{v \rm assign} \sim 20 {\kms}$ this leaves a residual uncertainty in the peculiar velocities of $\sigma_{v \rm res} \sim 80 {\kms}$, indicating that we reach about a factor of two more accurate estimates than previous works  \citep[see e.g.][based on linear theory]{Neill2007}.
\label{halo_disp}

We note that the peculiar velocity term in Eq.~\ref{eq:vsplit} is weighted with $1+z$ (see e.g., \citet[][]{Davis2011}) first within the reconstruction process  (\S \ref{sec:Constrained simulations}) and in the Hubble fit (see next section).

\section{Hubble constant measurements}
\label{sec:Results}

In this section we start presenting the input supernova sample used in our study, followed by our results on Hubble constant measurements applying the peculiar velocity correction shown in the previous section. We finally present our cosmic flows reconstructions and in particular our estimates on bulk flows and the Local Group motion.

\subsection{Input supernova data}
\label{sec:SNdata}

\begin{figure}
\centering
\includegraphics[width=.47\textwidth]
{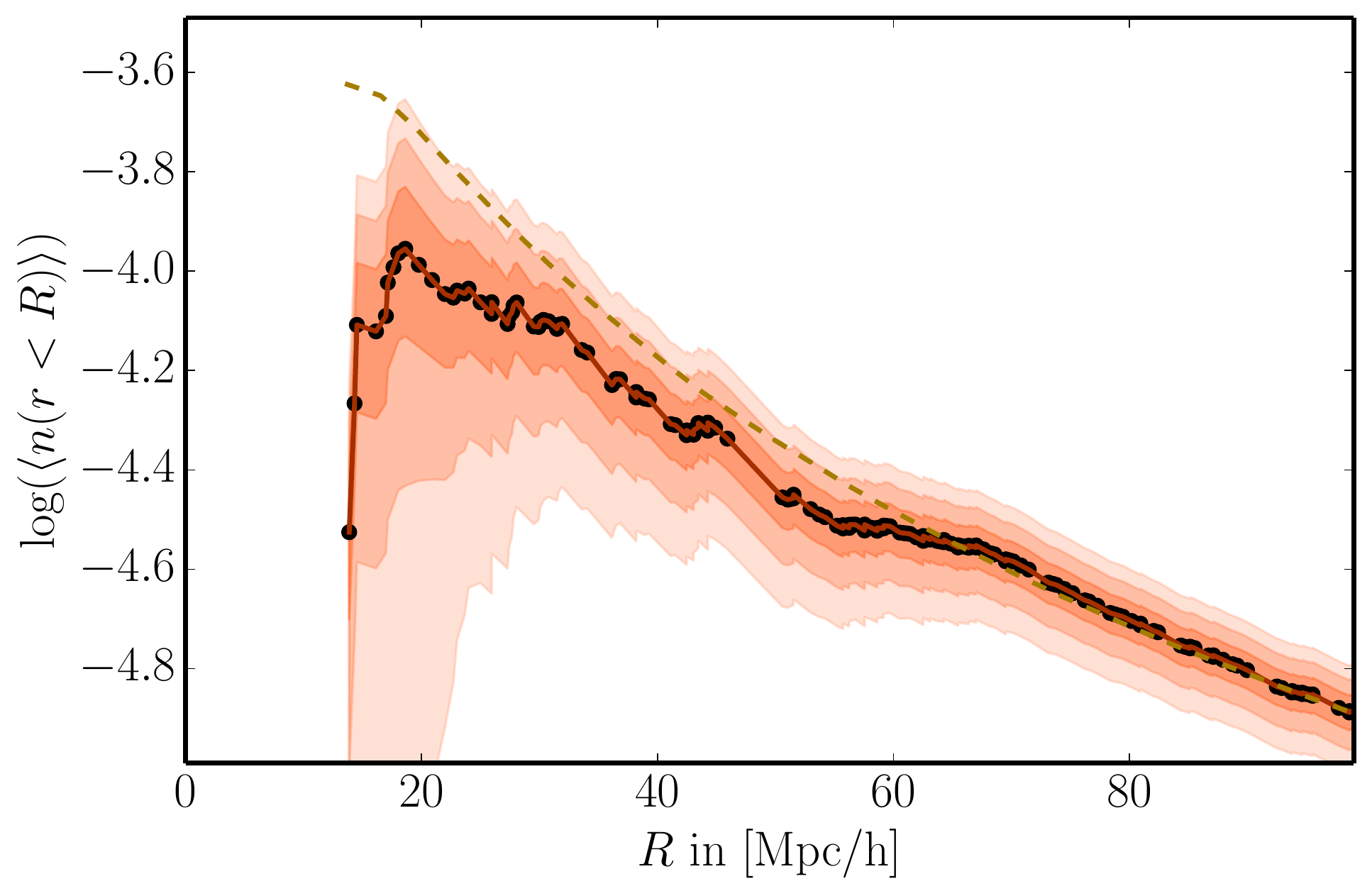}
\vspace{-2mm}

\caption{Cumulative supernova density as a function of luminosity distance $d_L$. Shades indicate $1,2,3\sigma$ poisson error. The dashed line indicates a slope of constant flux.}
\label{fig_SNdens}
\end{figure}

We use the data from the Extragalactic Distance Database (EDD) \footnote{publicly available at http://edd.ifa.hawaii.edu/dfirst.php}
\citep{EDD2009} comprising five sources \citep{Prieto2006,Jha2007,Hicken2009,Amanullah2010,Folatelli2010}, which was compiled by  \citep{Courtois2012}. It consists of 308 supernovae within $0<z<0.12$.
In particular we focus in this work on the local sample of 164 supernovae within $z<0.025$,  which is contained inside our simulated box.
 The density of supernovae at increasing distance is shown in Fig.~\ref{fig_SNdens}.

\begin{figure}
\centering
\includegraphics[width=.47\textwidth]
{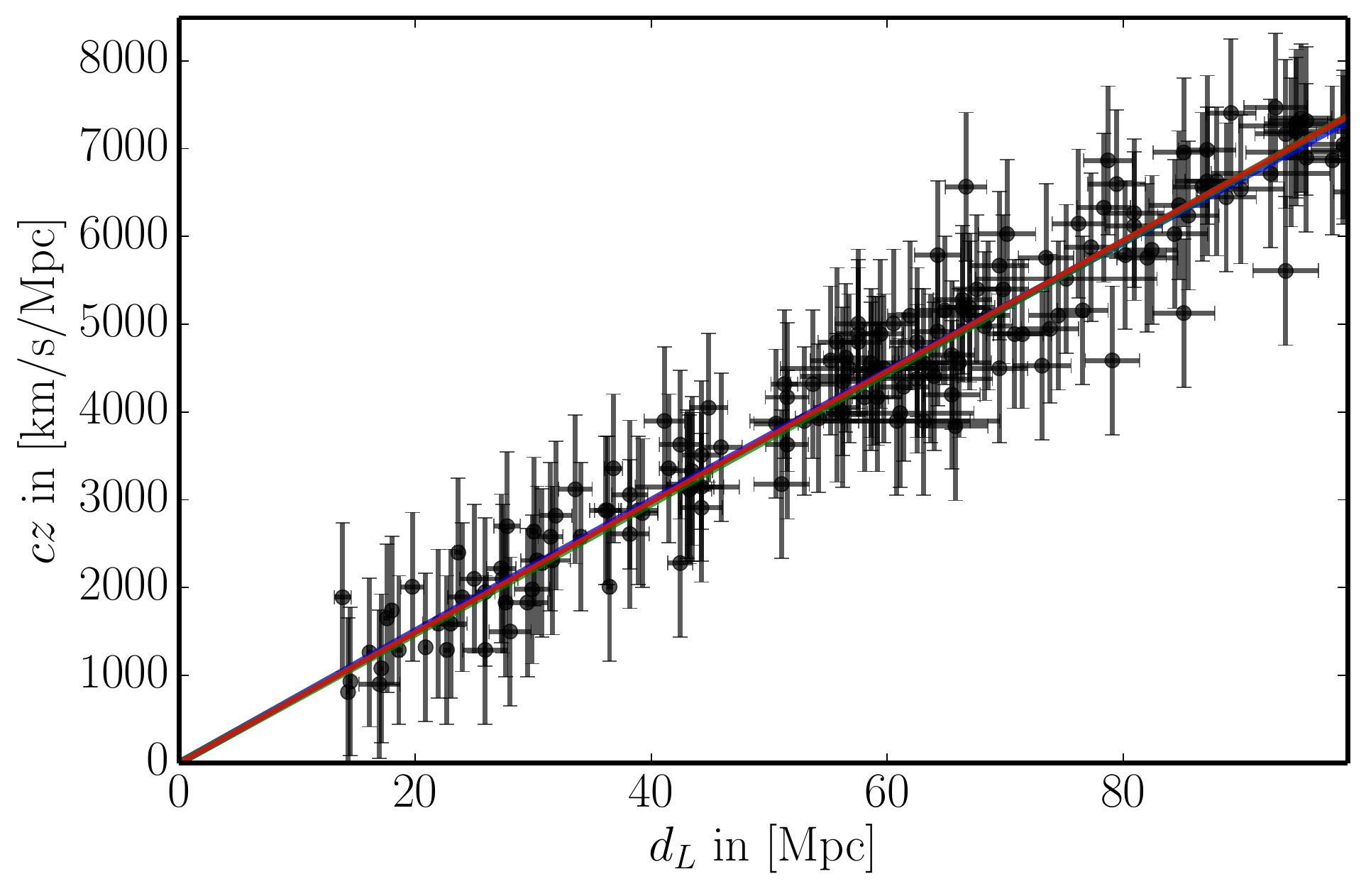}\\
\vspace{-5.5mm}
\includegraphics[width=.47\textwidth]
{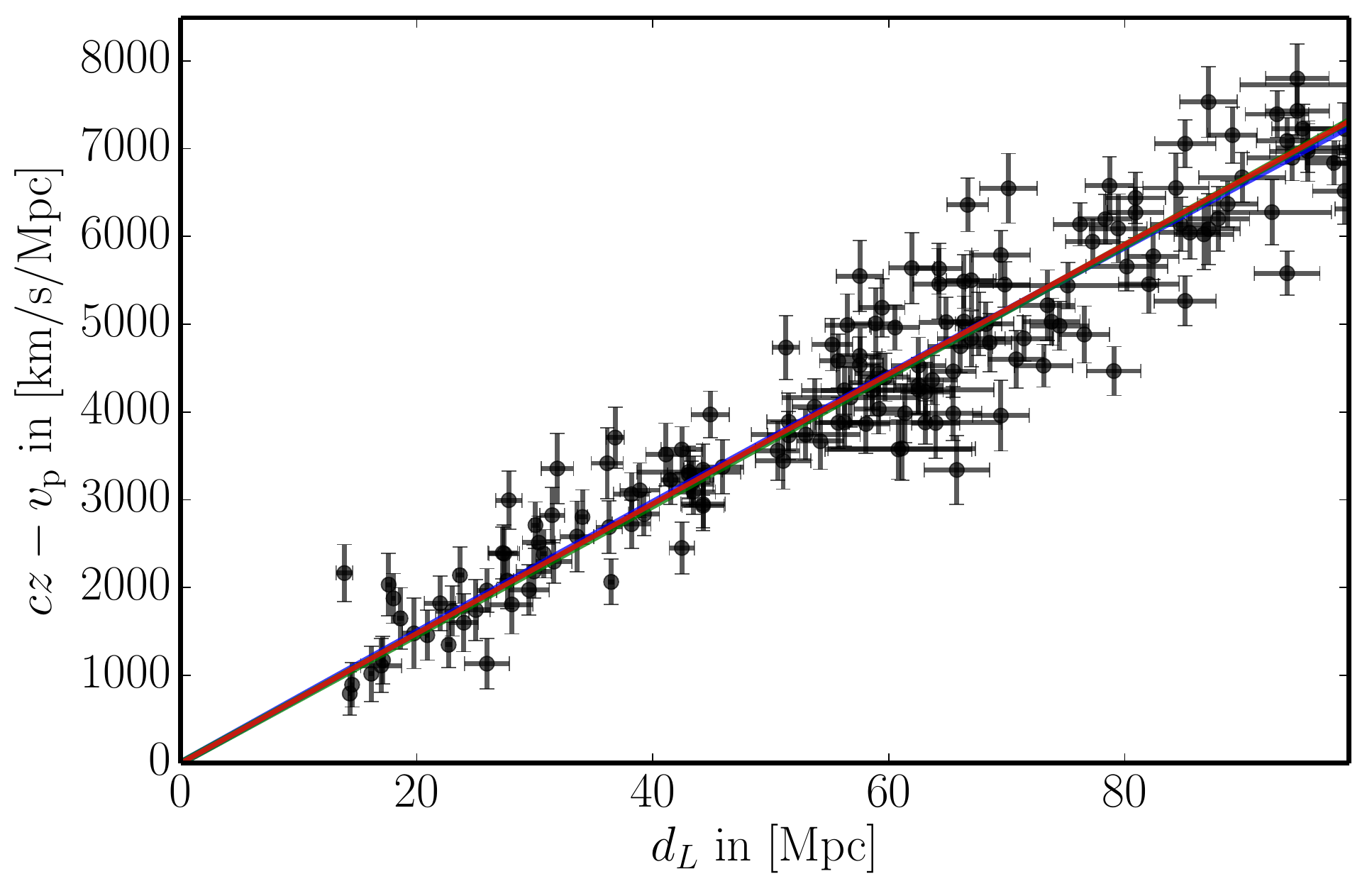}\\
\vspace{-2mm}

\caption{Recession velocity vs. Luminosity distance. Top panel: measured recession velocity. Bottom panel: measured recession velocity accounting for the influence of peculiar velocities. The solid lines show the Hubble constant fits,  red: assuming  a linear, green: a relativistic and blue: an approximate integral relation (see \S~\ref{sec:Distancecalculation}).}
\label{fig_Hubble}
\vspace{3mm}
\end{figure}

\subsection{Distance calculation}
\label{sec:Distancecalculation}

We proceed now to perform a linear regression of the supernova data based on their spectroscopic redshifts $cz$ and the corresponding inferred luminosity distances $d_{\rm L}$. 
Let us consider the cases  with and without redshift space distortions, i.e., $f(H_0d_{\rm L})=H_0d_{\rm L}-\mbi v_{\rm SN}\cdot\hat{\mbi r}$ and $f(H_0d_{\rm L})=H_0d_{\rm L}$, respectively, and their relation to the redshift distance $g(cz)$: $g(cz)=f(H_0d_{\rm L})$. In particular we consider following three approximations to study the robustness of our results:   linear relation $g(cz)=cz$,  relativistic relation $g(cz)=c\frac{2z+z^2}{2+2z+z^2}$ \citep[for a discussion see][]{DavisLineweaver2004}, and approximate integral relation  $g(cz)=cz+(1-\frac{3}{4}\Omega)cz^2+(9\Omega-10)\frac{\Omega}{8}cz^3$ \citep[see][]{Pen1999}. The first two are approximations for low redshifts \citep[for a discussion see][]{DavisLineweaver2004} independent of cosmology. The latter depends on the matter density $\Omega_{\rm M}$ and we fix  $\Omega_{\rm M} h^2$ according to WMAP7 \citep{wmap7}.
 In the case without peculiar motions corrections we assume a radial velocity dispersion of $\sigma_v=282 {\kms}$, as extracted from our analysis in \S \ref{halo_disp}. This accounts for the uncertainty due to peculiar velocities within the Chi-square fit. These error bars are considerably reduced when adding the actual information of the peculiar motions according to our analysis.

We find slightly different Hubble constant estimates depending on the choice of approximation ($g(cz)$), as we show in  Fig.~\ref{fig_Hubble}. We should also note that the calibration of the supernovae have a considerable impact on the absolute value ranging from about 74 to 76 ${\rm \kmsmpc}$ \citep[see e.g.][]{Courtois2012}. We note that recent environmental studies indicate that the impact could be even larger \citep[][]{Rigault2013,Rigault2014}.  However the relative shift (using the same approximation) due to the peculiar velocity correction remains unchanged within error bars. Therefore we will focus in this work on the relative corrections and not on the absolute estimates. In fact we find, that taking peculiar motions corrections into account yields a correction of $\Delta H_0 =\dHcosmo {\rm \kmsmpc}$ for $z<0.025$. Reducing the range to $0.01<z<0.025$, and hence excluding the local super-cluster, as is done in several studies \citep[see e.g.][]{Riess2009,BenDayan2014}, we find a correction of  $\Delta  H_0 =\dHcosmob {\rm \kmsmpc}$ (Jackknife Error Estimates). Hence, these results reduce the tension between local and more distant probes of $H_0$ by about 3\%.

\citet{Riess2009} report a change of $1.0-1.2\kmsmpc$ in their Hubble constant estimation if they include $0.01<0.023$ in their study. 

In agreement with \citet{Riess2009} we find a change of the correction of $-0.41 \kmsmpc$ if we include the $0.01<0.023$.

\section{Discussion}
\label{sec:dis}

In \S \ref{sec:divergence} we discussed different possible sources of systematic deviations in the measurement of the Hubble constant, which originate in the presence of peculiar motions (see Eq.~\ref{eq:hubble_split_terms}).

Let us analyse our results in this section and the causes of the biases in the Hubble constant measurement.

\begin{figure}
\raggedright
\includegraphics[width=.47\textwidth]
{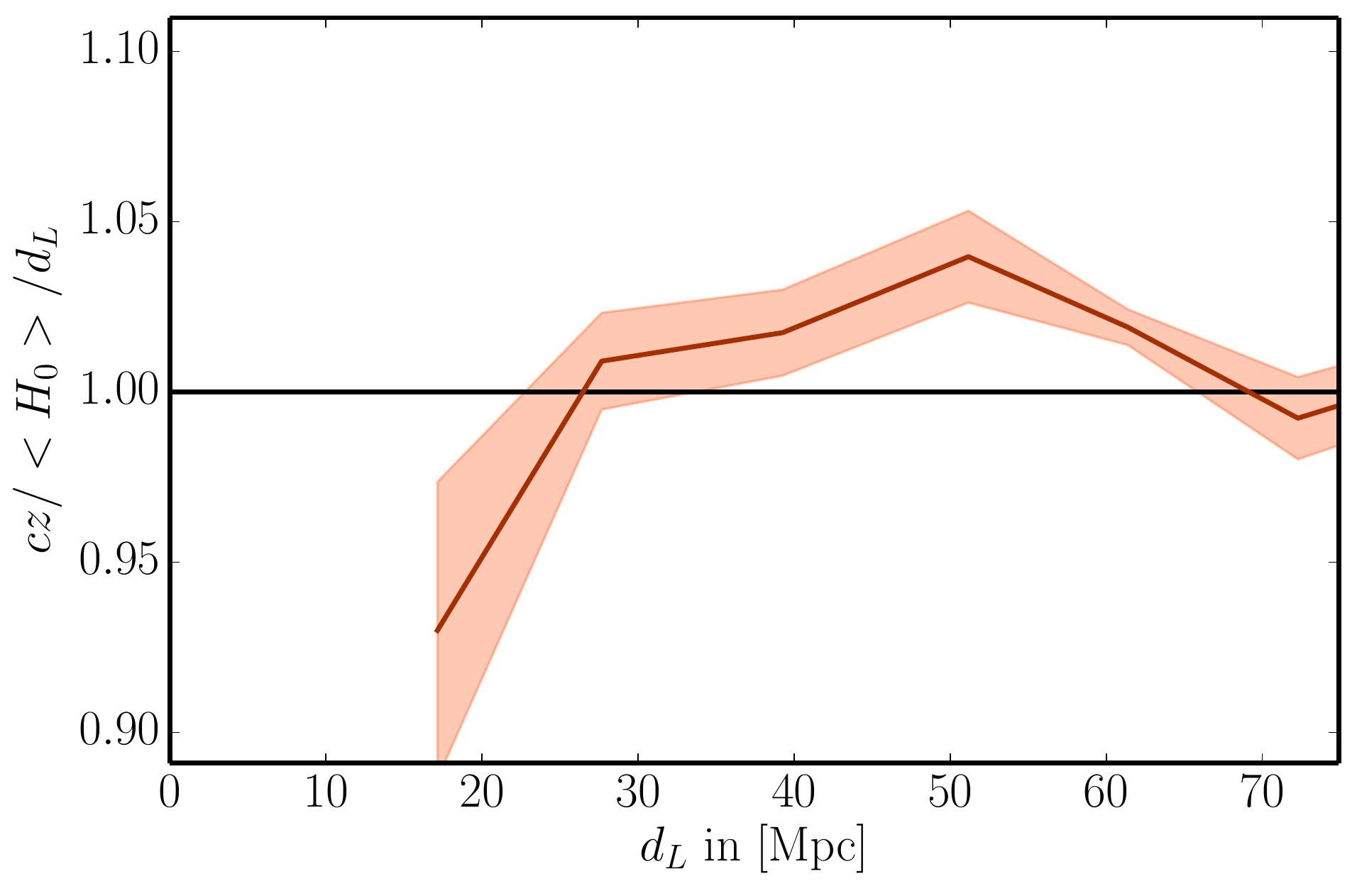}
\vspace{-2.75mm}
\vspace{-2.75mm}
\includegraphics[width=.47\textwidth]
{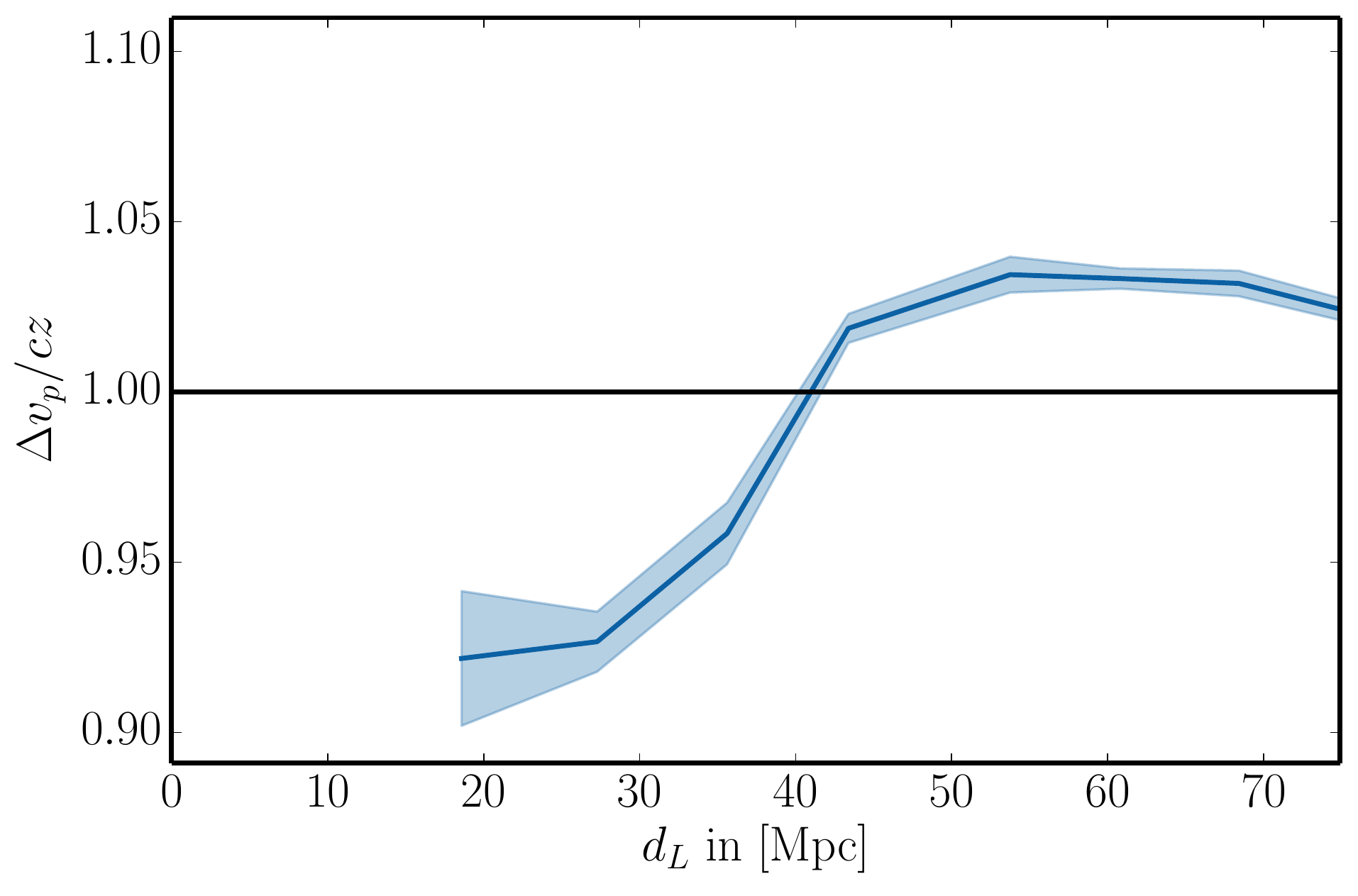}\\
\vspace{-1.5mm}

\caption{
Top panel: median of residuals from a  Hubble constant $\H0m_{\rm obs}$ fit (according to Eq.~\ref{eq:hubble}). Bottom panel: relative corrections as computed by constrained simulations. Shaded regions indicate relevant standard errors.}
\label{fig_dldrec}
\vspace{3.0mm}

\end{figure}

 The cumulative distribution of supernovae shows a deviation towards low distances from a flux limited sample following a one over squared distance law (see Fig.~\ref{fig_SNdens}). Beyond the radial selection effect and inhomogeneity, there is an apparent anisotropic distribution of supernovae. These effects have an impact on the Hubble measurement through the second and third terms in \eq{eq:hubble_split_terms}.

Fig.~\ref{fig_dldrec} indicates the residuals of a linear fit. There is a correspondence between the deviations and the corrections for distances that are well inside the reconstructed volume.

From these two figures it is apparent that there is a supernova under-density at radii smaller than $\sim 60 \Mpch$ leading to a radially diverging flow. According to \eq{eq:hubble_split_terms} this will then lead to positively biased Hubble constant estimations.  

\begin{figure}
\centering
\includegraphics[width=.475\textwidth]
{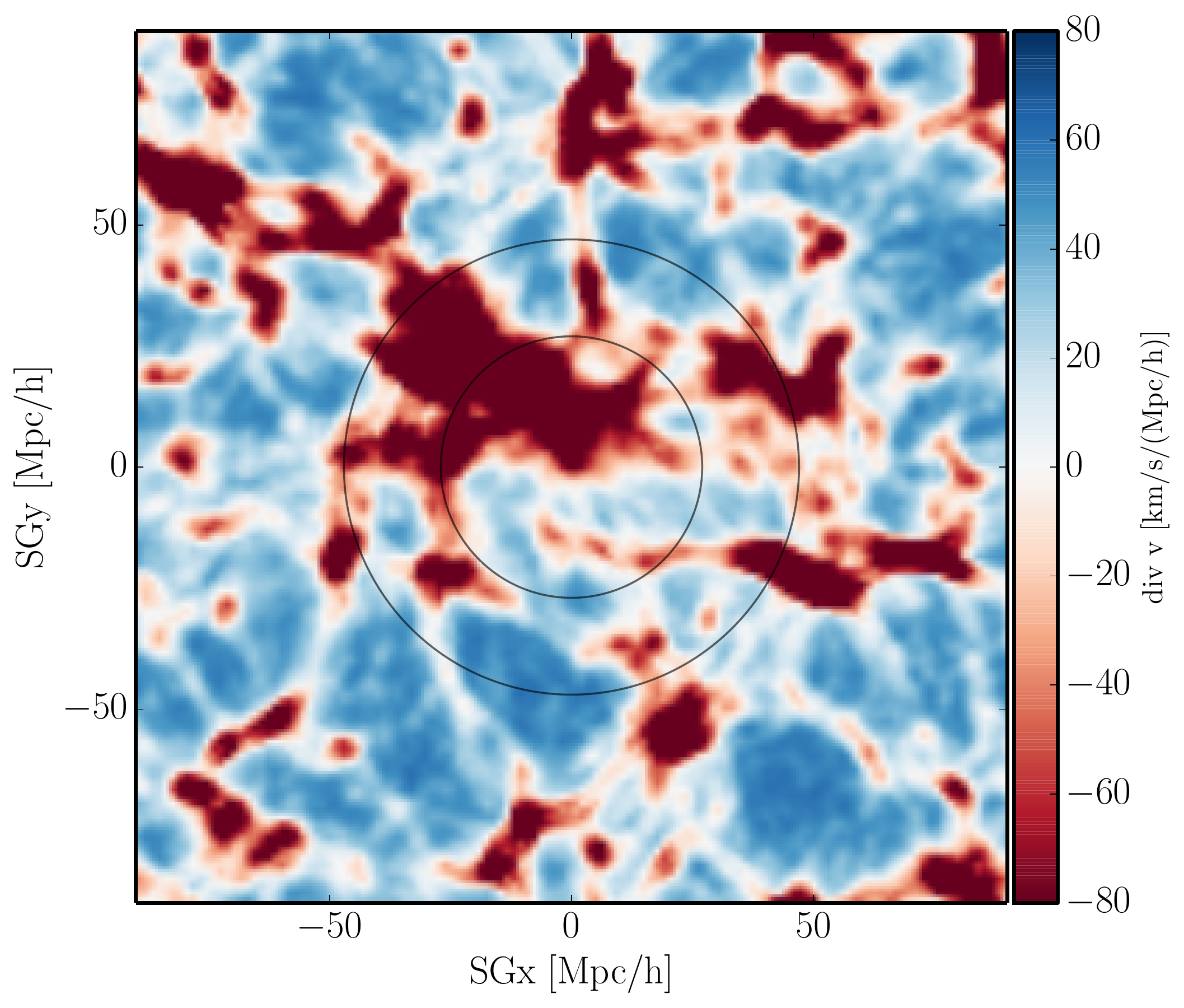}
\includegraphics[width=.475\textwidth]
{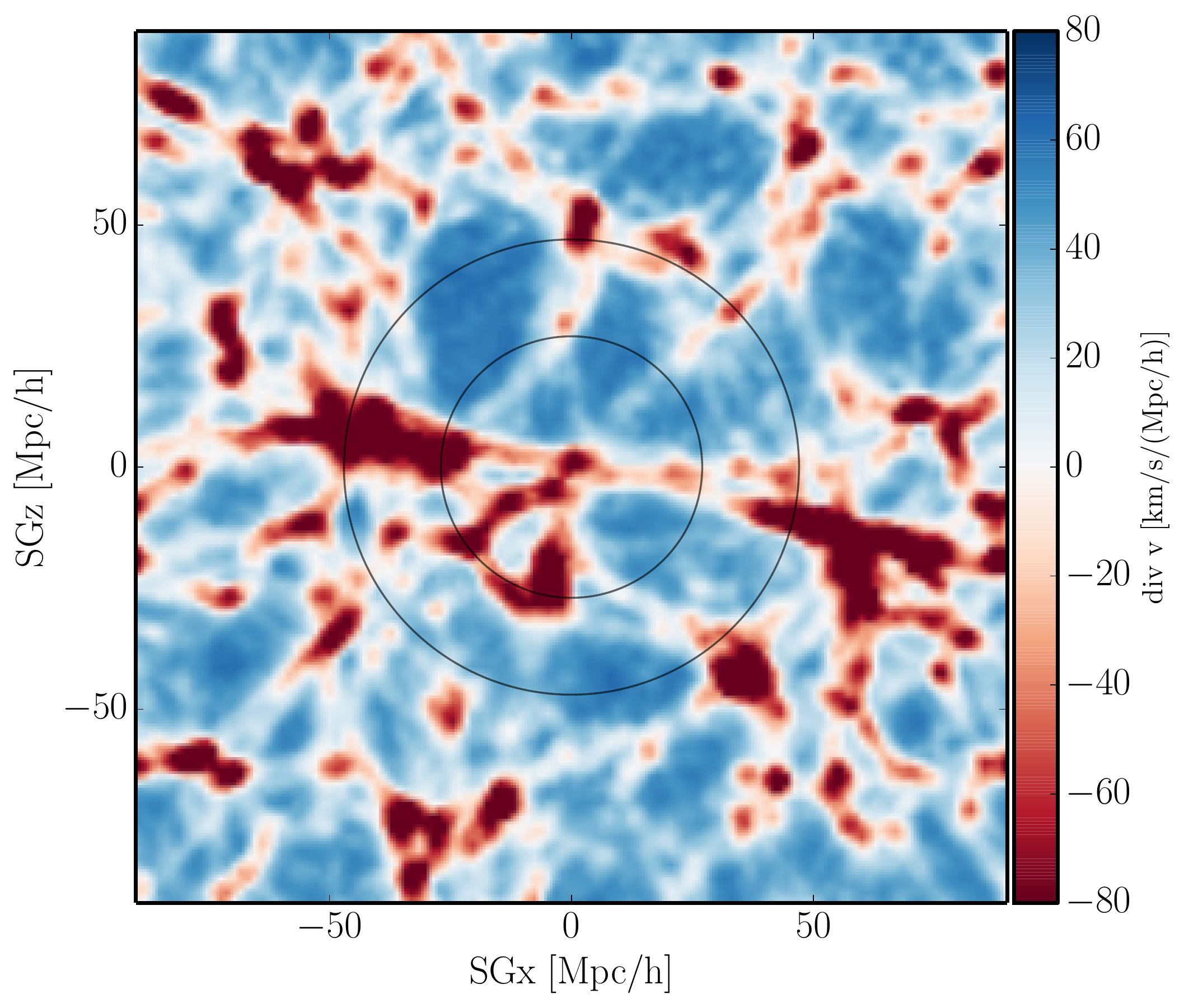}
\includegraphics[width=.475\textwidth]
{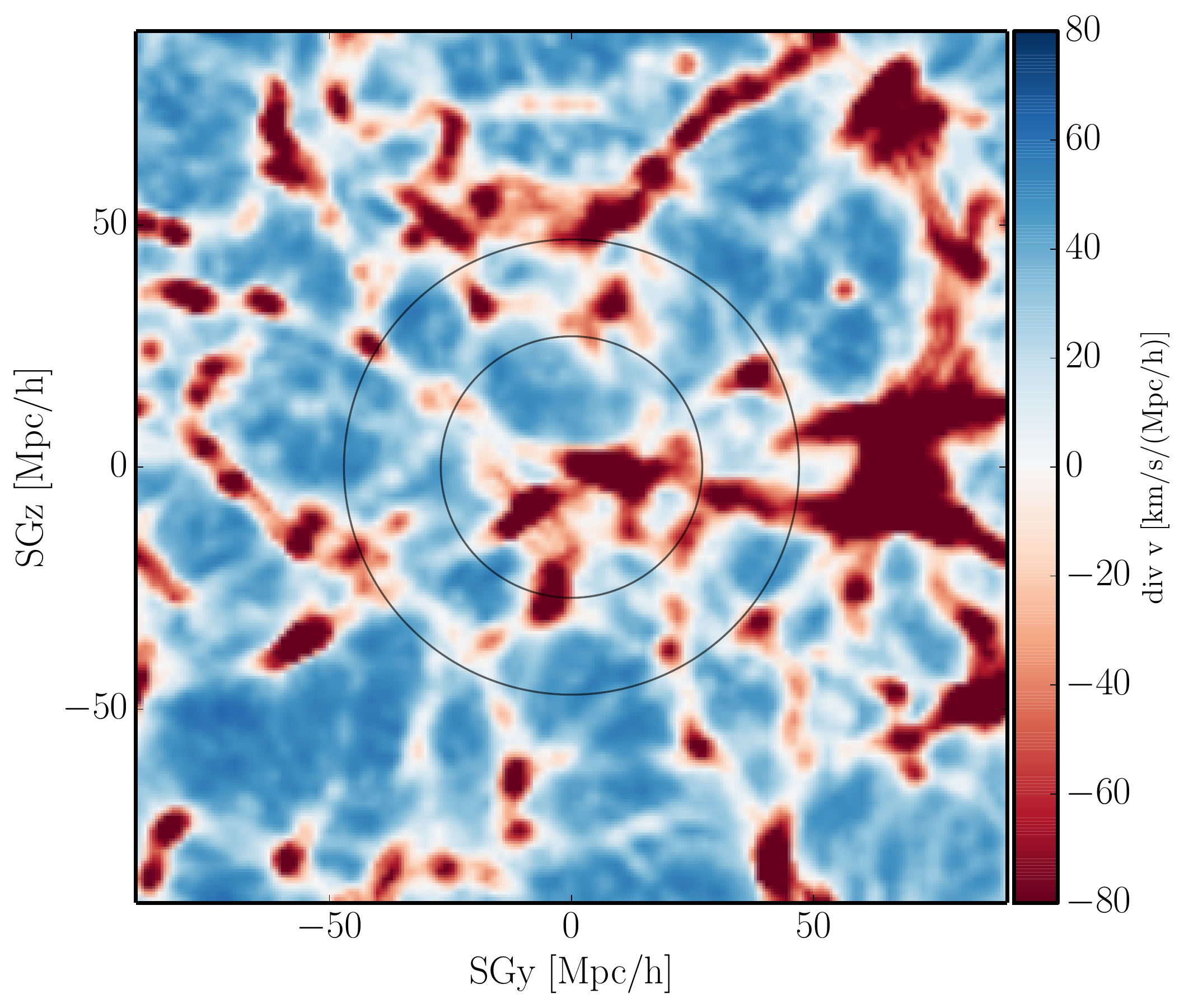}
\vspace{-4mm}
\caption{Maps of velocity divergence of a slice through an ensemble average of augmented constrained simulations (CS). The ensemble consists of 750 boxes - 25 CS, each of which has been rerun 30 times with random large scale modes. 
Collapsing regions are shown in red and diverging regions in blue. We note that between \moatinner and \moatouter (black circles) diverging (blue) regions dominate.
}
\label{fig_mapsmoat}
\end{figure}

Let us now analyse in detail the first term of Eq.~\ref{eq:hubble_split_terms}  dominated by  the velocity divergence.
Fig.~\ref{fig_mapsmoat} shows slices through the velocity divergence field. Here the constrained simulations have been augmented with random large scale modes and the average of the resulting velocity divergence field is shown.

\begin{figure}
\raggedright
\includegraphics[width=.47\textwidth]
{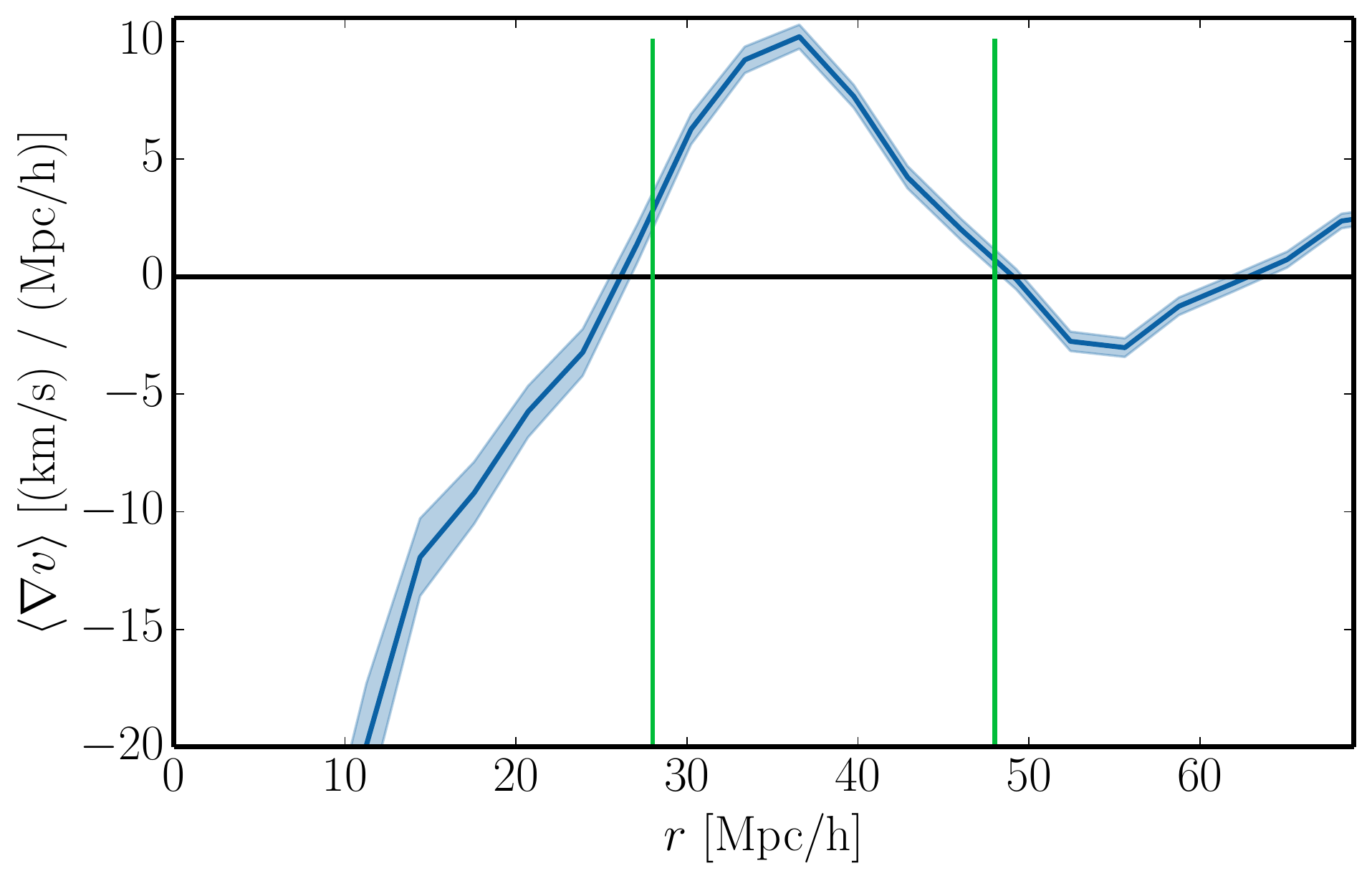}
\vspace{-5.5mm}
\includegraphics[width=.47\textwidth]
{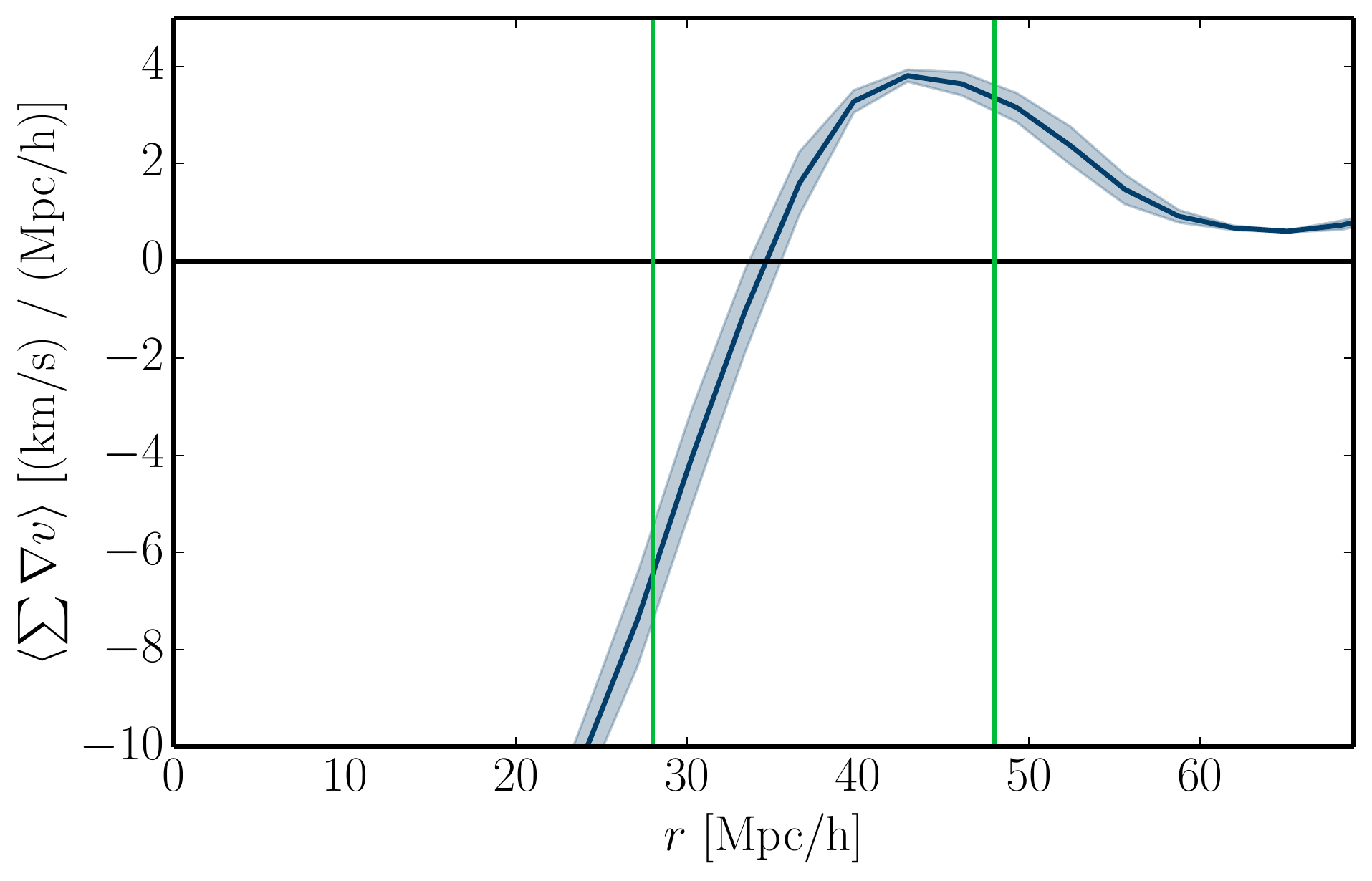}
\vspace{2.5mm}

\caption{Top: Velocity divergence measured in spherical bins with distance $r$. The shaded region indicates the standard deviation. Bottom: cumulative velocity divergence.
}
\label{fig:divv_vs_r}
\label{fig_divv_vs_r}
\end{figure}

It appears that in a shell between \moatinner and \moatouter there is an expanding region (enclosed by black circles in Fig.~\ref{fig_mapsmoat}). The velocity divergence field as a function of radius $r$ in Fig.~\ref{fig_divv_vs_r} clearly demonstrates the expansion at distances above \moatinner just beyond the Virgo Supercluster (VSC) (enclosed by green lines in Fig.~\ref{fig_divv_vs_r}). Under-densities in that region have been noted before, as it comprises e.g. the local void \citep{Tully2008}. Moreover, within supernova data a signal has been detected at $4800 {\kms}$ by \citet{Jha2007}. However we want to stress that this region is not a void by any standard definition since it is not a convex volume and contains considerable overdensities including part of the Great Attractor. Nevertheless it is a spherical shell that is on average diverging.

Effectively this diverging region acts like a ``Hubble bubble''  such as described by (\citet{Zehavi1998,Jha2007,Conley2007}, since this underdense region expands and leads to an apparently higher Hubble constant measurement.

The bottom panel of Fig.~\ref{fig_divv_vs_r} indicates the impact of the peculiar velocity field on each tracer, solely due to the velocity divergence term. 
We find that the velocity divergence is indeed the dominating correction by comparing the
 cumulative velocity divergence (bottom panel of Fig.~\ref{fig_divv_vs_r}) to the absolute peculiar velocity correction  $\Delta \vpec$ (bottom panel of  Fig.~\ref{fig_dldrec}).

\section{Conclusions}
\label{sec:Conclusions}

We have presented a detailed analysis of the Hubble constant measurement corrections in the Local Universe taking into account the impact of cosmic flows and density perturbations. Our findings indicate that low redshift supernova samples ($0.01<z<0.025$) overestimate the Hubble constant by about 3\%. By making the appropriate peculiar motion corrections we find that the Hubble constant is reduced by $\Delta  H_0 =\dHcosmob {\rm \kmsmpc}$ ($\Delta  H_0 =\dHcosmo {\rm \kmsmpc}$ for $z<0.025$). This correction should be considered in addition to environmental dependences which affect the systematic errors in the standardization of SNe Ia.
Due to the small enclosed volume and the small distances, local observations of recession velocities are easily influenced by inhomogeneities, anisotropies and peculiar velocity.
Our analysis relies on precise constrained \Nbody simulations which are based on unprecedented Bayesian self-consistent phase-space reconstructions of the initial conditions of the Local Universe using the 2MRS data.
This ensemble of constrained nonlinear velocity fields permits us to deal with cosmic variance, which is paramount to unlock the full potential of the abundant supernovae distance measurements in the Local Universe. 
We have furthermore accounted for periodicity effects and missing attractors by increasing the simulations volume up to distances of $360 \Mpch$, running 750 simulations with augmented Lagrangian perturbation theory.
 The cosmic flows reconstructions presented in this work show an extraordinary resemblance with independent observations. In particular we obtain consistent results to observed bulk flows on scales up to $40 \Mpch$ based on the Tully-Fisher relation with the 2MTF galaxies.  Furthermore, our calculations yield a  Local Group speed of $|v_{\rm LG}|=685\pm137$ s$^{-1}$km ($l=260.5\pm 13.3$, $b=39.1\pm 10.4$) compatible with the observed CMB dipole velocity.  Our analysis suggests that there is a missing component of about 130 $\kms$. All these results show a remarkable agreement with $\Lambda$CDM.

We investigate the origin of our peculiar velocity correction further and find two main components. The first is the correction due to the alignment of the large scale flow and the anisotropy of the supernovae distribution.
The second contribution is due to a diverging shell at distances from \moatinner to \moatouter. 
Despite the very mild deviation from average density within a sphere of $z<0.025$, this shell is below average density. It is located beyond the Virgo Super-cluster and limited by well known overdensities such as e.g. the Perseus-Pisces or the Coma clusters.

Still a number of aspects could be further improved in this work.
Constrained simulations with even higher precision would be able to reduce the remaining small scale uncertainties further. This involves the challenges of improving the structure formation model and the redshift space distortion treatment, including \Nbody solutions within the reconstruction process.

Reconstructions on larger volumes would allow us to pick up more extended inhomogeneities and flows that we included as uncertainties. However this requires a treatment of the Kaiser rocket effect and ever more sparse spectroscopic redshift observations on the full sky awaiting for new data.  

This work represents a first attempt to make a full nonlinear analysis of the local cosmic flows and the Hubble constant measurement from galaxy redshift data.

\section*{Acknowledgements}
The authors thank Uros~Seljak for suggesting this project and giving them valuable advice. SH thanks the Department of Physics, the Department of Astronomy, and the Lawrence Berkeley National Laboratory at the University of California for hospitality during the beginning of this work. The authors thank Brent~Tully, Enzo~Branchini, Radek~Wojtak, Stefan Gottl{\"o}ber and Yehuda~Hoffman for useful discussions.  SH acknowledges support by the Deutsche Forschungsgemeinschaft under the grant
$\mathrm{GO}563/21-1$. The constrained simulations have been performed at the Juelich Supercomputing Centre (JSC). Furthermore SH acknowledges the support from Matthias~Steinmetz.

\bibliography{paper}
\bibliographystyle{mn2e.bst}


\end{document}